# Evidence for Energy Supply by Active Region Spicules to the Solar Atmosphere

S. Zeighami[1,2] · A. R. Ahangarzadeh Maralani [1,2] · E. Tavabi [3] · A. Ajabshirizadeh[1,2]

**Abstract** We investigate the role of active region spicules in the mass balance of the solar wind and energy supply for heating the solar atmosphere. We use high cadence observations from the *Solar Optical Telescope* (SOT) onboard the *Hinode* satellite in the Ca II H line filter obtained on 26 January 2007. The observational technique provides the high spatio-temporal resolution required to detect fine structures such as spicules. We apply Fourier power spectrum and wavelet analysis to SOT/*Hinode* time series of an active region data to explore the existence of coherent intensity oscillations. The presence of coherent waves could be an evidence for energy transport to heat the solar atmosphere. Using time series, we measure the phase difference between two intensity profiles obtained at two different heights, which gives information about the phase difference between oscillations at those heights as a function of frequency. The results of a fast Fourier transform (FFT) show peaks in the power spectrum at frequencies in the range from 2 to 8 mHz at four different heights (above the limb), while the wavelet analysis indicate dominant frequencies similar to those of the Fourier power spectrum results. A coherency study indicates the presence of coherent oscillations at about 5.5 mHz (3 min). We measure mean phase speeds in the range $250 - 425 \, \text{km s}^{-1}$ increasing with height. The energy flux of these waves is estimated to be $F = 1.8 \times 10^6 - 11.2 \times 10^6 \, \text{erg cm}^{-2} \text{s}^{-1}$ or $1.8 - 11.2 \, \text{kW m}^{-2}$ which indicates that they are sufficiently energetic to accelerate the solar wind and heat the corona to temperatures of several million degrees. We compute the mass flux carried by spicules as $3 \times 10^{-10} - 2 \times 10^{-9} \, \text{g cm}^{-2} \text{s}^{-1}$, which is $10 - 60$ times larger than the mass

✉ S. Zeighami
 zeighami@iaut.ac.ir

✉ A. R. Ahangarzadeh Maralani
 ahangarzadeh@iaut.ac.ir

 E. Tavabi
 tavabi@iap.fr

 A. Ajabshirizadeh
 ali_ajabshir@yahoo.com

[1] Center for Excellence in Astronomy and Astrophysics (CEAA), Research Institute for Astronomy and Astrophysics of Maragha (RIAAM), Maragha, Iran

[2] Department of Physics, Tabriz Branch, Islamic Azad University, Tabriz, Iran

[3] Physics Department, Payame Noor University (PNU), 19395-3697-Tehran, I. R. of Iran

carried away from the corona because of the solar wind (around $3 \times 10^{-11} \mathrm{g\,cm^{-2} s^{-1}}$). Therefore, our results indicate that about $0.02 - 0.1$ of the spicule mass is ejected from the corona, while the rest goes down to the chromosphere. In other words, spicules can balance the mass loss due to the slow solar wind.

## 1. Introduction

The mechanisms responsible for solar coronal heating are still an unresolved problem. There have been several attempts to understand the energetics of the solar chromosphere and corona in recent years. Researchers have proposed that the key for solving this problem could be the understanding of the nature of the fine-scale transient events distributed all over the solar surface. Among these, spicules are the most prominent small-scale dynamic phenomena in chromospheric regions. Spicules are rapidly changing hair-like jets of relatively cool material ejected from the low chromosphere and protruding into the hot corona.

Spicules can heat the corona either by ejecting hot plasma into the corona (De Pontieu *et al.,* 2011) or by providing the energy dissipated by MHD waves (De Pontieu *et al.,* 2007b; He *et al.,* 2009b). Spicules are visible at the limb in rather cool chromospheric lines (Hα and Ca II). Dynamic structures are formed when photospheric oscillations and convective flows leak into the chromosphere along magnetic field lines, where they form shocks to drive jets of chromospheric plasma, as suggested by De Pontieu, Erdélyi, and James (2004). De Pontieu *et al.* (2007b) found that MHD waves are fairly ubiquitous in spicules. For an observational review on spicule oscillations see Zaqarashvili and Erdélyi (2009) and Tsiropoula *et al.* (2012). De Pontieu *et al.* (2007b) and He *et al.* (2009a) have detected kink or Alfvén waves that could carry enough energy to heat the corona.

De Pontieu *et al.* (2007a) identified two types of spicules. Type I spicules have typical lifetimes of $150 - 400 \mathrm{s}$. They show rising and falling motions following parabolic trajectories with maximum rise velocities from 15 to $40 \mathrm{\,km\,s^{-1}}$, often with a deceleration between 50 and $400 \mathrm{\,m\,s^{-2}}$, which is not equal to solar gravity. De Pontieu, Erdélyi, and James (2004); Hansteen *et al.* (2006), and De Pontieu *et al.* (2007b) suggested that dynamic fibrils are driven by magnetoacoustic shocks that result from flows and waves that leak along magnetic field lines from the photosphere into the chromosphere. Type I spicules´ behavior is similar to that of active region (AR) dynamic fibrils and some quiet Sun (QS) mottles (Suematsu, Wang, and Zirin, 1995) as observed in Hα (Hansteen *et al.,* 2006; De Pontieu *et al.,* 2007b, and Rouppe van der Voort *et al.,* 2007). These authors found a linear correlation between the deceleration and maximum velocities observed in fibrils and mottles. Advanced MHD simulations in two and three dimensions reproduce the observed parabolic paths, decelerations, and maximum velocities very well (Hansteen *et al.*, 2006; De Pontieu *et al.*, 2007b). The lifetime of these jets depends on the inclination of the magnetic field in the photosphere and low chromosphere. Inclined field decreases the acoustic cutoff frequency and allows leakage of the underlying photospheric wave spectrum with the dominant 5 min p-modes (Jefferies *et al.*, 2006; McIntosh and Jefferies, 2006). Regions where the field is more vertical are dominated by the chromospheric acoustic cutoff frequency corresponding to a 3 min period. Observations of type I spicules are difficult in coronal holes because of the inclination of the magnetic field. The magnetic field in coronal holes is mostly unipolar, so that the overall direction of the field is more vertical and leads to less leakage of p-modes, and therefore to less energetic type I spicules. This linear correlation can be readily understood in terms of shock-wave physics (De Pontieu *et al.*, 2007a; Heggland, De Pontieu, and Hansteen, 2007).

Type II spicules have only upward motions, shorter lifetimes, and higher velocities. Pereira, De Pontieu, and Carlsson (2012) determined typical velocities from 30 to $110\,\mathrm{km\,s^{-1}}$, and lifetimes from 50 to 150 s. Type II spicules are the type that mostly dominates both quiet Sun and coronal holes; but in active regions type I spicules dominate (De Pontieu *et al.*, 2007a; Pereira, De Pontieu, and Carlsson, 2012).

The interface between the Sun's photosphere and corona, *i.e.* the chromosphere and transition region, play a key role in the formation and acceleration of the solar wind. Tian *et al.* (2014) studied the presence of intermittent small-scale jets with speeds of $80-250\,\mathrm{km\,s^{-1}}$ from the narrow bright network lanes of the interface region observed with the *Interface Region Imaging Spectrograph* (IRIS). They found that the jets are likely an intermittent but persistent source of mass and energy for the solar wind. They suggested that some of these network jets are likely the on-disk counterpart and transition region manifestation of type II spicules (see also Pereira *et al.*, 2014).

He *et al.* (2009b) found that spicules are modulated by high-frequency ($\geq 0.02$ Hz) transverse fluctuations that propagate upward along them with phase speeds from 50 to $150\,\mathrm{km\,s^{-1}}$. They also observed that some of the modulated spicules show clear wave-like shapes with wavelengths of less than 8 Mm. Tavabi (2014) studied off-limb and on-disk spicules with the aim of finding a counterpart of limb spicules on the disk and found a definite signature with a strong power at a period of 3 min (5.5 mHz) and 5 min (3.5 mHz).

Tavabi *et al.* (2015b) analysed in detail the proper transverse motions of mature and tall polar-region spicules for different heights. Assuming that there might be helical-kink waves or Alfvénic waves propagating inside their multi-thread structure, they interpreted the quasi-coherent behavior of all visible components as presumably confined by a surrounding magnetic envelope. Tavabi *et al.* (2015a) studied coherency of solar spicules intensity oscillations. They using by FFT and wavelet analysis calculated dominant frequencies and phase speeds in QS, AR and active sun.

Ahangarzadeh Maralani *et al.* (2015) investigated quiet-Sun spicule oscillations estimated the dominant frequency peaks to be at 3.5 mHz, 5.5 mHz, and 8 mHz, and the dominant coherency frequency at about 3.5 mHz and 5.8 mHz; these results are a strong evidence for upwardly propagating coherent waves. They found mean phase speeds in the range $175-300\,\mathrm{km\,s^{-1}}$ which increased with height.

Sakao *et al.* (2007), using *Hinode's X-ray Telescope* observations at the edge of an active region in the neighborhood of a coronal hole, suggested that the observed outflows could be one of the sources of the slow solar wind. Harra *et al.* (2008) speculated that the material in the persistent outflows at the edges of an AR observed with the EUV *Imaging Spectrometer* (EIS) onboard *Hinode* could become part of the slow solar wind. They analyzed the magnetic field around the AR and concluded that the outflow speeds could reach values above $100\,\mathrm{km\,s^{-1}}$. Tian *et al.* (2012) by *Hinode/*EIS data found mainly two types of oscillations, the first type is found at loop footpoint regions and the second type is associated with the upper part of loops. They also found that the first type oscillations are likely to be signatures of quasi-periodic upflows and the second type oscillations are more likely to be signatures of kink/Alfvén waves rather than flows.

Kudoh and Shibata (1999) performed MHD simulations of torsional Alfvén waves propagating along an open magnetic flux tube in the solar atmosphere. They showed that the energy flux ($3\times10^{5}\,\mathrm{erg\,cm^{-2}\,s^{-1}}$) transported into the quiet corona is enough to heat this region.

Tsiropoula and Tziotziou (2004) investigated the role of chromospheric fine structures, *e.g.* mottles in the mass balance and heating of the solar atmosphere by studying two-dimensional high-resolution H$\alpha$ observations.

Tavabi *et al.* (2015a) obtained dominant frequencies and phase speed of oscillations. They focused on comparing FFT and wavelet analyses. In the present work, we analyse the same observations of them but concentrate on priority and role of AR spicules, in energy and ass supply, with high processing more than those performed by Tavabi *et al.* (2015a). We obtain the frequency of oscillations using two different methods, FFT and wavelet analysis at different heights, and investigate the existent of coherent oscillations, and the phase speed between two heights using relative phase differences to find energy and mass flux, which was not discussed at all in Tavabi *et al.* (2015a). It is well known that the fast solar wind originates from coronal holes, while the slow solar wind may originate from active region boundaries. AR spicules certainly have nothing to do with the fast solar wind but may be associated with type II chromospheric spicules (De Pontieu *et al.,* 2007c) or transition region network jets (Tian *et al.,* 2014). In this article, we analyze the AR borders (see Sakao *et al.,* 2007; Harra *et al.,* 2008). We investigate the evidence for mass balance and energy supply of AR spicules to the solar atmosphere.

## 2. Observations

We use time sequences of AR data taken at the Sun´s equatorial limb observed in the Ca II H line with the SOT/*Hinode* (the wavelength pass-band is centered at 396.8 nm with a FWHM of 0.3 nm). The characteristics of the observations are summarized in Table 1.

**Table 1** The datasets obtained with the SOT/*Hinode* in the Ca II H line

| Date | Start and end times (U.T.) | X-center Y-center (arcsec) | Cadence (s) | Size (pixels$^2$) | X-FOV Y-FOV (arcsec) | pixel size (arcsec) |
|---|---|---|---|---|---|---|
| 26 January 2007 | 06:16:10 07:11:58 | 966.188 −71.258 | 8 | 1024×1024 | 55.7875 55.7875 | 0.0541 (∼40 km) |

We use the standard SOT subroutines for calibration of raw (level 0) data. These subroutines correct the CCD readout anomalies, bad pixels, flat-field, and subtract the dark pedestal. The SOT routine FG_PREP reduces the image spikes, jitter, and aligns the time series.

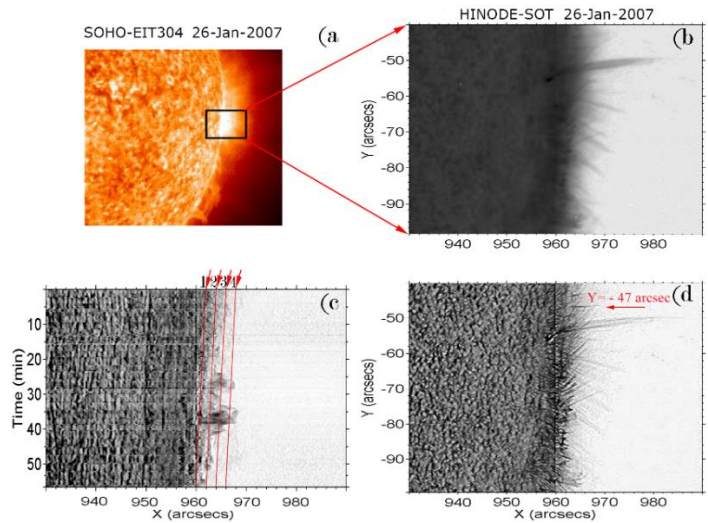

**Figure 1** (a) A section of an *EUV Imaging Telescope* (EIT), onboard the *Solar and Heliospheric Observatory* (SOHO), full disk image on 26 January 2007 where the AR is well seen. (b) An example of a negative image obtained with SOT in Ca II H (396.8 nm) on 26 January 2007 near the Equator. (c) An example of a negative time-X diagram (for Y= -47 arcsec). The positions of four different heights, which were selected for analysis are shown by four red color arrows (1, 2, 3, and 4) at a distance 35 pixels from each other. A drift speed toward the East was identified. (d) An example of a highly processed negative image to improve the visibility of the fine features after applying the Mad-Max operator.

## 3. Data Analysis

Using the Mad-Max operator we reduce considerably the background noise with high processing more than those performed by Tavabi *et al.* (2015a), (Koutchmy and koutchmy, 1989; November and koutchmy, 1996; Tavabi, Koutchmy, and Ajabshirizadeh, 2013; Tavabi, 2014) as shown in Figure 1d. We prepare time-slices of the processed images in the Y direction to study the spatial variations of the spicules at different times. The algorithm locates a specific row from all the images (Tavabi, 2014), see Figure 1. We analyze all pixels in the Y direction. Figure 1c shows only one example (for Y= - 47 arcsec).

We plot the intensity profiles in function of time at four separate heights increasing from the solar limb with a separation of 35 pixels between each other, respectively (*i.e.* reaching a maximum height equivalent to 4200 km above the limb). The four separate heights are shown in Figure 1c with red color arrows labeled as 1, 2, 3, and 4. We use a FFT analysis to obtain the power spectra of the intensity profiles. An example of the results can be seen in Figure 2. The plots show peaks in the range from 2 to 8 mHz at the four selected heights. We note that the results are obtained for selected time slices and that they are consistent with those obtained by Tavabi (2014) and Lawrence and Cadavid (2012).

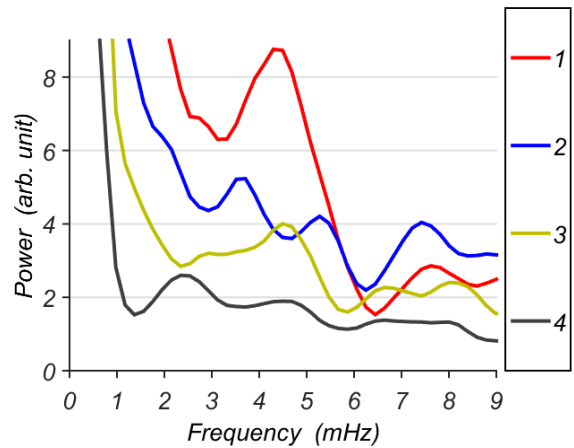

**Figure 2** An example of Fourier power spectra corresponding to four different heights labeled by four red color arrows in Figure 1 (corresponding to Y= - 47 arcsec).

Figure 3 shows an example of the 2D and 3D wavelet results which are consistent with the FFT results. Zaqarashvili and Erdélyi (2009), Gupta *et al*. (2013), and Tavabi *et al.* (2015a) have obtained similar results with only 2D representations. Bloomfield *et al*. (2004) applied the FFT transform and wavelet phase coherence analysis to a weak solar magnetic network region and showed the advantage of the wavelet analysis over more classical techniques, such as the Fourier analysis.

We use cross-spectrum phase and coherence estimates and the cross-wavelet transform software package (Torrence and Compo, 1998) between two heights and calculate the coherency, phase difference, and cross-wavelet oscillations (Figure 4; see 2D representation in Tavabi *et al*., 2015a). The results show coherency or a frequency of about 5.5 mHz (3 min).

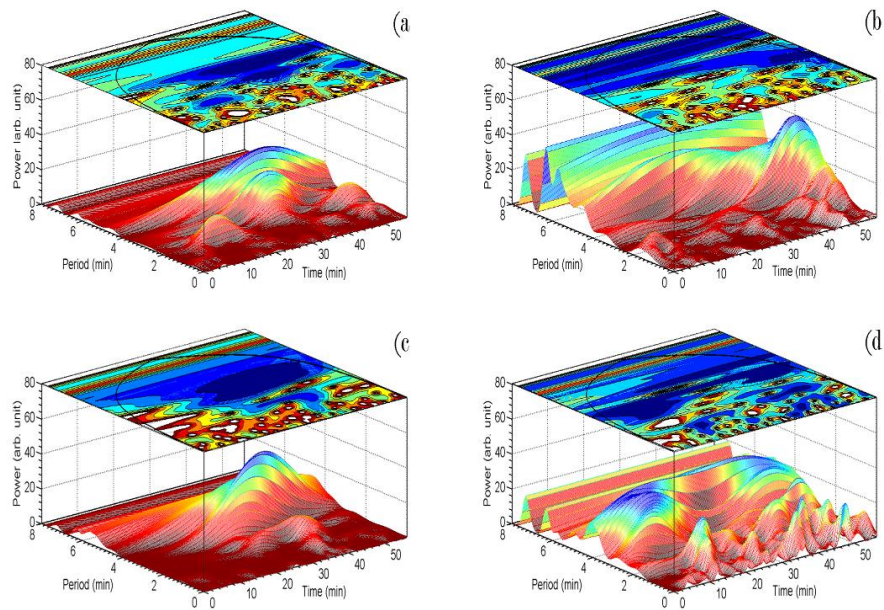

**Figure 3** (a) − (d) represent an example of the 2D and 3D wavelet results obtained from intensity profiles corresponding to the four heights that are shown by four red color arrows in Figure 1c ( corresponding to Y= - 47 arcsec). The solid black curve in the plot at the top of each panel indicates the cone of influence region (COI) where the wavelet power spectra are distorted because of the influence of the end points of finite length signals.

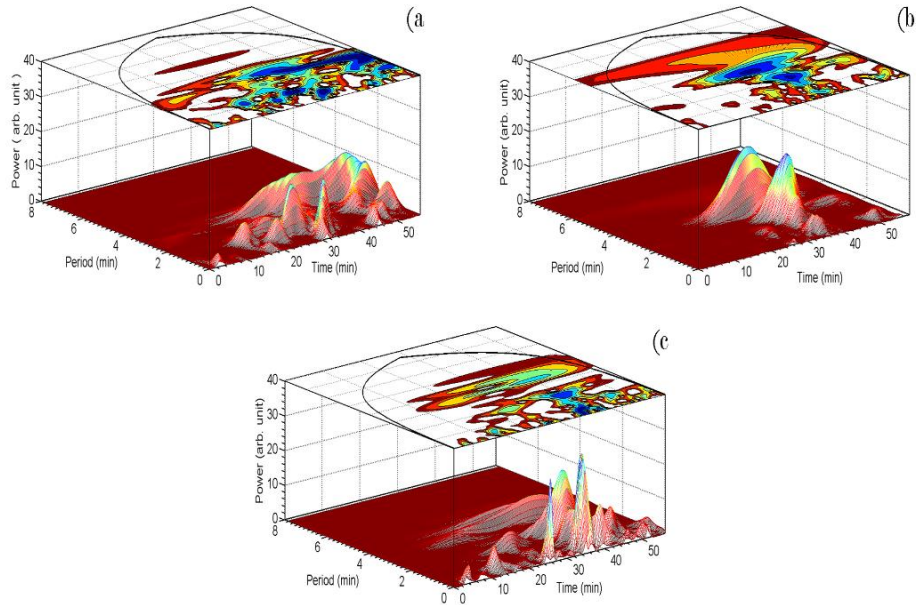

**Figure 4** An example of wavelet power-spectrum results for coherency between two heights. (a) − (c) show the wavelet coherency spectrum between two heights, *i.e.* first and second, second and third, third and fourth height, respectively (corresponding to Y= - 47 arcsec). The black curve in the plot at the top of each panel has a similar meaning as in Figure 3.

Measurements of the phase difference between two intensity profiles obtained at the two heights yields information on the phase delay between oscillations as a function of frequency (Trauth, 2007; Oppenheim and Verghese, 2010). These calculations are done for maximum coherency (coherency above a 75% level). The results of the phase difference analysis present coherency at frequencies similar to those of the cross-wavelet results (Figure 5).

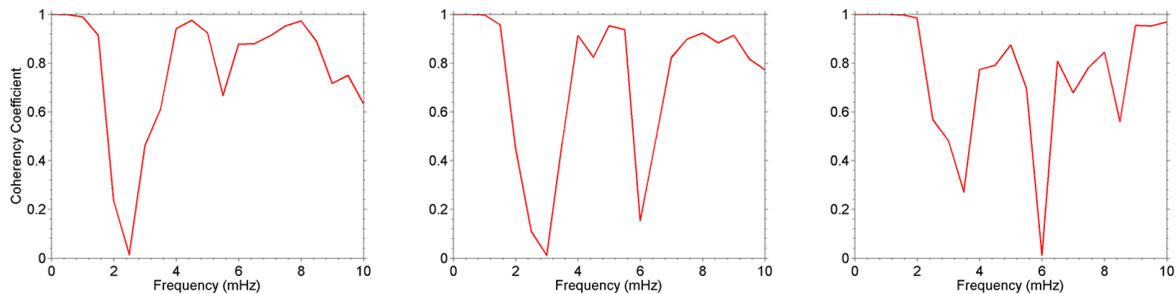

**Figure 5** An example of coherency plots. The plots show the coherency of waves propagating between different heights. The left, middle and right panels show coherency between two heights, *i.e.* first and second, second and third, third and fourth height, respectively (corresponding to Y= - 47 arcsec).

Using the equation for the phase difference as a function of frequency, $\Delta\varphi = 2\pi f\, T$ where $f$ is the frequency (in Hz) and $T$ is the time difference (in seconds), we obtain the time difference between two waves propagating between two heights and then we calculate the phase speed between these heights using the equation $V_{ph} = H/T$, where H is the distance between the two heights. We calculate the phase difference and phase speed for maximum coherency from all of coherency plots. In this way, we take into account all the time-slice results. Mean phase speeds between two heights are $250\,\text{km}\,\text{s}^{-1}$, $350\,\text{km}\,\text{s}^{-1}$ and $425\,\text{km}\,\text{s}^{-1}$, respectively. The phase speeds increase with height.

According to the energy flux equation, $F = \frac{1}{2}\rho v^2 V_A$ (Sterling, 2000; Ofman and Wang, 2008; McIntosh *et al.*, 2011), where $\rho$, $v$, $V_A$ are the density, wave amplitude, and Alfvén speed, respectively, we can estimate the energy flux of these waves. We take the Alfvén speed in the range $V_A = 250 - 425\,\text{km}\,\text{s}^{-1}$, the wave amplitude as $v = 20 - 30\,\text{km}\,\text{s}^{-1}$ ( Tavabi *et al.*, 2015b for type I), and the density as $\rho = 6.24 \times 10^{-11} - 40.4 \times 10^{-11}\,\text{kg}\,\text{m}^{-3}$ ( Tian *et al.*, 2014). We estimate the energy flux to be in the range $F = 1.8 \times 10^6 - 11.2 \times 10^6\,\text{erg}\,\text{cm}^{-2}\text{s}^{-1}$ or $1.8 - 11.2\,\text{kW}\,\text{m}^{-2}$ (based on Alfvénic waves), which indicates they are sufficiently energetic to accelerate the solar wind and heat the localized corona to multi-million degree. The mass flux carried by spicules depends on the fraction of the solar disk covered by them, $f$, the axial upward velocity, $v_u$, and the mass density, $\rho$. If we assume that half of the spicules show upward motions and that half of the material flows upward, the value of the mass flux carried upward is $F_m = \frac{1}{2}(\frac{1}{2}f\,\rho v_u)$ (Tsiropoula and Tziotziou, 2004). If we consider $f = 0.01$, as there are about 300,000 active spicules at any one time on the Sun's chromosphere, amounting to about 1% of the Sun's surface (see Freedman and Kaufmann, 2008) and $v_u = 20 - 30\,\text{km}\,\text{s}^{-1}$ (Tavabi *et al.*, 2015b), then $F_m = 3 \times 10^{-10} - 2 \times 10^{-9}\,\text{g}\,\text{cm}^{-2}\text{s}^{-1}$, which is $10 - 66$ times greater than the material flowing upward from the corona because of solar wind that is equal to $3 \times 10^{-11}\,\text{g}\,\text{cm}^{-2}\text{s}^{-1}$ (Ulmschneider, 1971). Therefore, our results show that about $0.02 - 0.1$ of the spicule material is ejected from corona and the rest goes down to the chromosphere. In other words spicules can balance the mass loss due to the solar wind. Tian *et al.* (2014) found that the mass loss rate is about 2−4 times larger than the total mass loss rate by the solar wind if, eventually, all the mass from the network jets is lost.

**4. Discussion**

The results of the Fourier power-spectrum plots show peaks at frequencies in the range from 2 to 8 mHz, which corresponds to a period of $2-8$ min at the four analyzed heights. It can be seen that the powers decrease with height. These periods are actually similar to the periods of intermittent AR outflows/upflows (*e.g.*, Nishizuka and Hara, 2011; Ugarte-Urra and Warren, 2011; Tian *et al.,* 2011,2012; De Moortel and Browning, 2015). Nishizuka and Hara (2011) analyzed the observations of an outflow region at the edge of an AR obtained with *Hinode*/EIS and found both continuous outflows and waves, which propagate from the base of the outflow. We suggest that these active regions outflows/upflows may represent the heated and accelerated spicules.

The wavelet analysis results indicate frequencies similar to the FFT results. The results of coherency and cross-wavelet studies between two heights show coherent oscillations with similar frequencies at about 5.5 mHz (3 min). Mean phase speeds obtained between two heights are $250\,\mathrm{km\,s^{-1}}$, $350\,\mathrm{km\,s^{-1}}$ and $425\,\mathrm{km\,s^{-1}}$, between the first and second, second and third, and third and fourth heights, respectively. The phase speed increases significantly with height along the spicules. This could be caused by a combination of two factors. First, at greater heights the Alfvén speed is higher since the plasma density is lower. Considering that the magnetic field strength is expected to decrease with height, the range of phase speeds at heights below 15" seems reasonable (Okamoto and De Pontieu, 2011; Tavabi *et al.*, 2015a; see also Ahangarzadeh Maralani *et al.*, 2015).

McIntosh *et al.* (2011) measured the phase speed of coronal-hole, quiet-Sun and AR disturbances (propagating continuous transverse motions that are associated with spicules) by cross-correlating the space–time plots at different heights above the limb. They determined a phase speed for AR disturbances of the order of $600\,\mathrm{km\,s^{-1}}$ with a small variance ($\sim 50\,\mathrm{km\,s^{-1}}$) observed along the loop structures using the *Atmospheric Imaging Assembly* on board the *Solar Dynamics Observatory* satellite (SDO/AIA) 171 Å channel. McIntosh *et al.* (2011) determined the period of oscillations in ARs to be $100-400$ s. Comparing the results of period and phase speed obtained by McIntosh *et al*. (2011) with our calculations they are of the same order.

Lawrence and Cadavid (2012) studied *Hinode/*SOT-FG observations of intensity fluctuations in the Ca II H line and the G-band and analyzed the intensity oscillation spectra considering relative phase differences, time delays, and cross coherences. They found that non-magnetic intensity fluctuations show a strong oscillatory power in the $3-7$ mHz band centered at 4.5 mHz, but this is suppressed as the magnetic field increases; they found that the maximum coherence is in the range $4-7$ mHz, which is consistent with our obtained values. Krishna Prasad, Banerjee, and Jagdev (2012) searched for oscillations in AR fan loops using simultaneous high cadence spectroscopic *Hinode/*EIS and SDO/AIA images and found two locations showing oscillations with short (<1 min to 3 min) and long periods ( ~ 9 min) using the wavelet technique. They detected outward propagating the disturbances with apparent propagation speeds from $83.5 \pm 1.8$ $\text{km s}^{-1}$ to $100.5 \pm 4.2 \text{ km s}^{-1}$ in 171 Å and 193 Å bands, respectively. Sekse *et al.* (2013) found oscillations that sometimes appear to propagate along the spicules with speeds between $18 - 108$ $\text{km s}^{-1}$ and can be interpreted as Alfvén waves.

There have been different views about the interpretation of periodic transverse motions as Alfvén waves. Van Doorsselaere, Nakariakov, and Verwichte (2008) studied waves in a stable cylindrical structure with a straight magnetic field. They named "Alfvén waves" to torsional axisymmetric MHD waves and "kink" waves to fast magnetosonic kink waves when structured by a cylindrical waveguide. Many authors have demonstrated the ubiquitous nature of Alfvén waves in magnetic flux tubes in a range of chromospheric and coronal plasmas (De Pontieu *et al.*, 2007b; Tomczyk *et al.*, 2007; Jess *et al.*, 2009). Van Doorsselaere, Nakariakov, and Verwichte (2008) proposed that these waves are better called fast mode kink waves rather than Alfvén waves. However, Goossens *et al.* (2009) considered these waves as surface Alfvén waves or Alfvénic waves. McIntosh *et al.* (2011) discussed and compared the kink and Alfvén waves. They pointed out that in highly dynamic, non-uniform plasmas such as the solar atmosphere, the MHD waves have mixed (*i.e.* slow, fast, Alfvén) properties depending on the local plasma parameters. It means that Alfvén waves are not just limited to "axisymmetric waves in a stable cylinder with a straight magnetic field" as suggested by Van Doorsselaere, Nakariakov, and Verwichte (2008).

McIntosh *et al.* (2011) also pointed out that the term Alfvén wave is commonly used by the fusion plasma physics, magnetospheric physics, solar wind physics, and space physics communities for a largely incompressible transverse wave for which the major restoring force is the magnetic tension. Tavabi *et al.* (2015b) investigated 2D and 3D images of spicules. According to the 2D images, they found that the waves had kink mode characteristics, but according to the 3D images, the wave propagation in a flux tube was seen as helical and kink, which they called helical-kink because the spicule axis was displaced. We follow Goossens *et al.* (2009), McIntosh *et al.* (2011), Tian *et al.* (2012), and Tavabi *et al.* (2015b) and use the word "Alfvénic".

We estimate the energy flux as $F = 1.8 \times 10^6 - 11.2 \times 10^6 \text{ erg cm}^{-2}\text{s}^{-1}$ or $1.8 - 11.2$ kW m$^{-2}$ (considering Alfvénic waves). Transverse motions of spicules are caused by Alfvénic waves. Preliminary estimates suggest that these Alfvénic waves carry an energy flux that may be of importance for the local energy balance and that, once they reach the corona, can play a significant role in the heating of the solar atmosphere and acceleration of the solar wind (De Pontieu *et al.*, 2007c).

Aschwanden (2001) found that the energy flux required for heating ARs, the quiet Sun and coronal holes are $2 \times 10^2 - 2 \times 10^3 \, \text{W m}^{-2}$, $10 - 2 \times 10^2 \, \text{W m}^{-2}$ and $5 - 10 \, \text{W m}^{-2}$, respectively. Therefore, our estimated energy flux could heat the solar corona. De Pontieu *et al.* (2007c) estimated the energy flux in the chromospheric Ca II H line as $4 \times 10^3 - 7 \times 10^3 \, \text{W m}^{-2}$. Their estimate is based on the observed velocity amplitude v ∼ $20 \, \text{km s}^{-1}$, conservative values for the spicule mass density ($\rho = 2.2 \times 10^{-11} - 4 \times 10^{-10}$ kg m$^{-3}$) with the implied Alfvén speeds in the order of $45 - 200 \, \text{km s}^{-1}$, which are compatible with our results.

Ofman and Wang (2008), using the Wentzel-Kramers-Brillouin (WKB) approximation, calculated the energy flux as $7 \times 10^5 - 5 \times 10^6 \, \text{erg cm}^{-2} \text{s}^{-1}$ which can balance typical radiative and conductive losses of an AR coronal loop and heat it to coronal temperatures.

He *et al.* (2009a) found that kink waves can carry an initial energy flux density of $1.5 \times 10^7 \, \text{erg cm}^{-2} \text{s}^{-1}$ in the low solar chromosphere, which decreases to $0.25 \times 10^6 \, \text{erg cm}^{-2} \text{s}^{-1}$ in the transition region. Therefore, there results show that the energy flux density carried by kink waves, despite its attenuation in the chromosphere and the transition region, is still sufficient for heating the quiet corona and/or driving the solar wind, which needs an energy input of the order of $10^6 \, \text{erg cm}^{-2} \text{s}^{-1}$.

Yang *et al.* (2015) estimated the energy flux to be ∼ $7 \times 10^6 \, \text{erg cm}^{-2} \text{s}^{-1}$ in simulations with Alfvén phase velocities ∼ $1000 \, \text{km s}^{-1}$. Kudoh and Shibata (1999) performed MHD simulations of torsional Alfvén waves propagating along an open magnetic flux tube in the solar atmosphere. They showed that the enough energy flux to heat the quiet corona (∼ $3 \times 10^5 \, \text{erg cm}^{-2} \text{s}^{-1}$) is transported into the corona. Our results of the estimated energy flux indicate that spicules are energetic enough to accelerate the solar wind and heat the localized corona.

The fast solar wind, flow speeds of $400 - 800 \, \text{km s}^{-1}$, emerges from magnetically open coronal holes in the quiet Sun, while the slow solar wind, with flow speeds of $250 - 400 \, \text{km s}^{-1}$, originates from above the more active regions on the Sun (Schwenn, 2000; Koutchmy, 2000). Okamoto and De Pontieu (2011) investigated the statistical properties of Alfvénic waves along spicules in a polar coronal hole using observations of *Hinode/*SOT and showed that the observed high-frequency waves carry a significant energy flux of the order of $2.5 \times 10^5 \, \text{erg cm}^{-2} \text{s}^{-1}$, in principle enough to play a role in heating the corona. McIntosh *et al.* (2011) determined energy flux densities in a coronal hole and the quiet corona of the order of $100-200\ \text{W}\,\text{m}^{-2}$ at a height of 15 Mm. Their study was based on the SDO/AIA observations of transition region and coronal emission and concluded that it is sufficient to provide the energy necessary to drive the fast solar wind. They also estimated the energy flux of low-frequency Alfvénic waves in denser active region loops as around $100\ \text{W}\,\text{m}^{-2}$, this is not sufficient to provide the $2000\ \text{W}\,\text{m}^{-2}$ required to power the active corona. Tian *et al.* (2014) obtained an energy flux of $4-24\ \text{kW}\,\text{m}^{-2}$ from coronal hole observations; they also found that the energy flux that reaches the corona is enough to drive the solar wind ($\sim 700\ \text{W}\,\text{m}^{-2}$).

Comparing our results with those of other authors mentioned above, we see that our estimated energy flux is greater than their calculated values. Active regions on the solar surface contain magnetic structures that have even higher mass densities and magnetic field strengths, in addition to a larger number of spicules. Therefore, the energy flux available to heat the corona will be significantly higher than the minimum value required for localized heating.

Sakao *et al.* (2007) suggested that the material flowing out from the edges of an AR *per* unit time is $\sim 2.8\times 10^{11}\ \text{g}\,\text{s}^{-1}$, which is equivalent to $\sim 0.25$ of the total solar wind mass loss rate ($1\times 10^{12}\ \text{g}\,\text{s}^{-1}$). Tsiropoula and Tziotziou (2004) analyzed a solar region containing several mottles, which constitute the fine structures of the quiet solar chromosphere, using high resolution Hα observations and obtained a globally averaged mass flux carried upward of about $7.1\times 10^{-9}\ \text{g}\,\text{cm}^{-2}\,\text{s}^{-1}$, which is two orders of magnitude larger than the outward flow of mass from the corona due to the solar wind. Tian *et al.* (2014) studied the solar wind sources (coronal holes) and found a mass loss rate about $2-24$ times larger than the total mass loss rate of the solar wind if, eventually, all the mass in the network jets is lost. We determine the value of the mass flux, $F_m$ as $3\times 10^{-10}-2\times 10^{-9}\ \text{g}\,\text{cm}^{-2}\,\text{s}^{-1}$ which is $10-60$ times larger than the coronal upward mass flux due to the solar wind. Therefore, our results indicate that about $0.02-0.1$ of the spicule material is ejected from the corona and the rest flows down to the chromosphere. In other words spicules can balance the mass losses due to the solar wind.

## 5. Conclusions

We have analyzed AR spicules above the solar limb using data taken with *Hinode*/SOT We have used Fourier power spectrum and wavelet analysis of time series data of an AR and found dominant frequencies in the range of $2-8$ mHz at four different heights above the limb. Results of coherency analysis indicate coherent oscillations at about 3.5 mHz. We have determined phase speeds in the range $250-425$ $\mathrm{km\,s^{-1}}$ that increased with height. The energy flux corresponding to the waves is estimated to be $F=1.8\times10^6 - 11.2\times10^6$ $\mathrm{erg\,cm^{-2}\,s^{-1}}$ or $1.8-11.2$ $\mathrm{kW\,m^{-2}}$. We have obtained the mass flux carried by spicules as $3\times10^{-10} - 2\times10^{-9}$ $\mathrm{g\,cm^{-2}\,s^{-1}}$ which is $10-60$ times larger than the material ejected upward from the corona ($3\times10^{-11}\,\mathrm{g\,cm^{-2}\,s^{-1}}$). Therefore, our results indicate that about $0.02-0.1$ of the spicule mass is ejected from the corona, and the rest flows down to the chromosphere. We conclude that AR spicules are sufficiently energetic to accelerate the solar wind and heat the corona to multi-million degree temperature and can balance the mass losses due to the slow solar wind. Of course there are some uncertainties in the flux estimates, for example the birthrate of these spicules, which is related to their number on the solar surface at a given time. This number depends on the spatial resolution, instrument and observation point of view, and of several other factors. Using different data sets obtained from other satellites and different lines of sight, we are able to show that for a range of AR characteristics, spicules can significantly play a role in heating the solar atmosphere.

**Acknowledgements:** *Hinode* is a Japanese mission developed and launched by ISAS/JAXA, with NAOJ as domestic partner and NASA and STFC (UK) as international partners. Image processing MAD-MAX program was provided by O. Koutchmy, see http://www.ann. jussieu.fr/~koutchmy/ index_newE.html and wavelet analysis software was provided by Torrence and Compo (http://atoc.Colorado.edu/research/wavelets). The authors are indebted to Tavabi *et al.* (2015a) for the helpful article and study of the paper can be useful for better understanding our paper. We warmly acknowledge the work of Editor and unknown Referee, who provided an extended detailed report and added many interesting suggestions and comments. This work has been supported by the Research Institute for Astronomy and Astrophysics of Maragha (RIAAM) and the Center for Excellence in Astronomy and Astrophysics (CEAA), and the Isalmic Azad University, Tabriz Branch (IAUT).

**Disclosure of Potential Conflicts of Interest**

The authors declare that they have no conflicts of interest.

**REFERENCES**

Ahangarzadeh Maralani, A.R., Zeighami, S., Tavabi, E., Ajabshirizadeh, A.: 2015, New Astron. submitted.

Aschwanden, M. J.: 2001, *Astrophys. J.* **560**, 1035. DOI: 10.1086/323064.

Bloomfield, D. S., McAteer, R. T. J., Lites, B.W., Judge, P. G., Mathioudakis, M., Keenan, F. P.: 2004, *Astrophys. J.* **617**, 623. DOI: 10.1086/425300.

De Moortel, I., Browning, P.: 2015, *Phil. Trans. R. Soc.* A *373*, 40269. DOI: 10.1098/rsta.2014.0269.

De Pontieu, B., Erdélyi, R., James, S. P.: 2004, *Nature* **430**, 536. DOI: 10.1038/nature02749.

De Pontieu, B., Hansteen, V.H., Rouppe van der Voort, L., van Noort, M., Carlsson, M.: 2007b, Astron. Soc. Pac. **368**, 65

De Pontieu, B., McIntosh, S.W., Carlsson, M., Hansteen, V.H., Tarbell, T.D., Boerner, P., *et al.*: 2011, *Science* **331**, 55. DOI: 10.1126/science.1197738.

De Pontieu, B., McIntosh, S.W., Carlsson, M., Hansteen, V.H., Tarbell, T.D., Schrijver, C.J., *et al.*: 2007c, Science 318, 1574. DOI: 10.1126/science.1151747.

De Pontieu, B., McIntosh, S., Hansteen, V.H., Carlsson, M., Schrijver, C.J., Tarbell, T.D., *et al.*: 2007a, Publ. Astron. Soc. Jap. 59, 655. DOI: 10.1093/pasj/59.sp3.S699.

Freedman, R.A., Kaufmann, W.J.: 2008, *Universe. New York, USA*: W. H. Freeman and Company, 762.

Goossens, M., Terradas, J., Andries, J., Arregui, I., Ballester, J. L.: 2009, *Astron.Astrophys*. **503**, 213. DOI: 10.1051/0004-6361/200912399.

Gupta, G.R., Subramanian, S., Banerjee, D., Madjarska, M.S., Doyle, J.G.: 2013, *Solar Phys.* **282**, 67. DOI: 10.1007/s11207-012-0146-y.

Hansteen, V. H., De Pontieu, B., Rouppe van der Voort, L., van Noort, M., Carlsson, M.: 2006, *Astrophy. J. Lett.* **647**, L73. DOI: 10.1086/507452.

Harra, L. K., Sakao, T., Mandrini, C.H., Hara, H., Imada, S., Young, P.R., *et al.*: 2008, *Astrophy. J.* **677**, L159. DOI: 10.1086/588288.

He, J., Marsch, E., Tu, C., Tian, H.: 2009a, *Astrophy. J.* **705**, L217. DOI: 10.1088/0004-637X/705/2/L217.

He, J.S., Tu, C.Y., Marsch, E., Guo, L.J., Yao, S., Tian, H.: 2009b, *Astron. Astrophys*. **497**, 525. DOI:10.1051/0004-6361/200810777.

Heggland, L., De Pontieu, B., Hansteen, V. H.: 2007, *Astrophy. J.* **666**, 1277. DOI: 10.1086/518828.

Jefferies, S.M., McIntosh, S.W., Armstrong, J.D., Bogdan, T.J., Cacciani, A., Fleck, B.: 2006, *Astrophy. J.* **648**, L151. DOI: 10.1086/508165.

Jess, D.B., Mathioudakis, M., Erdélyi, R., Crockett, P. J., Keenan, F. P., Christian, D. J.: 2009, *Science* **323**, 1582. DOI: 10.1126/science.1168680.

Koutchmy, S.: 2000, In: Murdin, P. (ed.) Encyclopedia of Astronomy and Astrophysics, Bristol, 2271. DOI: 10.1888/0333750888/2271.

Koutchmy, O., Koutchmy, S.: 1989, In: von der Lühe, O. (ed.) High Spatial Resolution Solar Observations, Proc. 10th Sacramento Peak Summer Workshop, National Solar Observatory, **217**.

Krishna Prasad, S., Banerjee, D., Jagdev S.: 2012, *Solar Phys*. **281**, 67. DOI: 10.1007/s11207-012-0098-2.

Kudoh, T., Shibata, K.: 1999, *Astrophys*. *J.* **514**, 493. DOI: 10.1086/306930.

Lawrence, J.K., Cadavid, A.C.: 2012, *Solar Phys.* **280**, 125. DOI: 10.1007/s11207-012-0059-9.

McIntosh, S., W., De Pontieu, B., Carlsson, M., Hansteen, V., Boerner, P., Goossens, M.: 2011, *Nature*, **475**, 477. DOI: 10.1038/nature10235.

McIntosh, S.W., Jefferies, S.M.: 2006, *Astrophys. J.* **647**, L77. DOI: 10.1086/507425.

Nishizuka, N., Hara, H.: 2011, *Astrophys. J.* **737**, L43. DOI: 10.1088/2041-8205/737/2/L43.

November, L. J., Koutchmy, S.: 1996, *Astrophys. J.* **466**, 512. DOI: 10.1086/177528.

Ofman, L., Wang, T. j.: 2008, *Astron. Astrophys.* **482**, L9. DOI: 10.1051/0004-6361:20079340.

Okamoto, T., De Pontieu, B.: 2011, *Astrophys. J.* **736**, L24. DOI: 10.1088/2041-8205/736/2/L24.

Oppenheim, A.V., Verghese, G.C.: 2010, Introduction to Communication, Control and Signal Processing, Massachusetts, Class Notes for 6.011.

Pereira, T. M. D., De Pontieu, B., Carlsson, M.: 2012, *Astrophys. J.* **759**, 18. DOI: 10.1088/0004-637X/759/1/18.

Pereira, T.M.D., De Pontieu, B., Carlsson, M.; Hansteen, V., Tarbell, T. D., Lemen, J., *et al.:* 2014, *Astrophys. J.* **792**, 15. DOI: 10.088/2041-8205/792/1/L15.

Rouppe van der Voort, L., De Pontieu, B., Hansteen, V., Carlsson, M., van Noort, M.: 2007, *Astrophys. J.* **660**, L169. DOI: 10.1086/518246.

Sakao, T., Kano, R., Narukage, N., Kotoku, J., Bando, T., DeLuca, E.E., *et al.*: 2007, *Science* **318,** 1585. DOI: 10.1126/science.1147292.

Schwenn, R.: 2000, In: Murdin, P. (ed.) Encyclopedia of Astronomy and Astrophysics, Bristol, 2301. DOI: 10.1888/0333750888/2301.

Sekse, D. H., Rouppe van der Voort, L., De Pontieu, B., Scullion, E.: 2013, *Astrophys. J.* **769**, 44. DOI: 10.1088/0004-637X/769/1/44.

Sterling, A. C.: 2000, *Solar Phys.* **196**, 79. DOI: 10.1023/A: 1005213923962.

Suematsu, Y., Wang, H., Zirin, H.: 1995, *Astrophys. J.* **450**, 411. DOI: 10.1086/176151.

Tavabi, E.: 2014, Astrophys. Space Sci. 352, 43. DOI: 10.1007/s10509-014-1901-3.

Tavabi, E., Ajabshirizadeh, A., Ahangarzadeh Maralani, A.R., Zeighami, S.: 2015a, J. Astrophys. Astron. 36, 307. DOI: 10.1007/s12036-015-9335-z.

Tavabi, E., Koutchmy, S., Ajabshirizadeh, A.: 2013, *Solar Phys.* **283**,187. DOI: 10.1007/s11207-012-0011-z.

Tavabi, E., Koutchmy, S., Ajabshirizadeh, A., Ahangarzadeh Maralani, A.R., Zeighami, S.: 2015b, *Astron. Astrophys.* **573**, A4. DOI: 10.1051/0004-6361/201423385.

Tian, H., DeLuca, E. E., Cranmer, S. R., De Pontieu, B., Peter, H., Martínez-Sykora, J., *et al.*: 2014, *Science* **346**, 1255711. DOI: 10.1126/science.1255711.

Tian, H., McIntosh, S.W., Habbal, S.R., He, J.: 2011, *Astrophys. J.* **736**, 130. DOI: 10.1088/0004-637X/736/2/130.

Tian, H., McIntosh, S.W., Wang, T., Ofman, L., De Pontieu, B., Innes, D.E., *et al.*: 2012, *Astrophys. J.* **759**, 144. DOI: 10.1088/0004-637X/759/2/144.

Tomczyk, S., McIntosh, S.W., Keil, S.L., Judge, P.G., Schad, T., Seeley, D.H., *et al.*: 2007, *Science* **317**, 1192. DOI: 10.1126/science.1143304.

Torrence, C., Compo, G. P.: 1998, *Bull. Am. Met. Soc.* **79**, 61.

Trauth, M. H.: 2007, MATLAB Recipes for Earth Sciences, Springer, Berlin.

Tsiropoula, G., Tziotziou, K.: 2004, Astron. Astrophys. 424, 279. DOI: 10.1051/0004-6361:20035794.

Tsiropoula, G., Tziotziou, K., Kontogiannis, I., Madjarska, M. S., Doyle, J. G., Suematsu, Y.: 2012, *Space Sci. Rev.* **169**, 181. DOI: 10.1007/s11214-012-9920-2.

Ugarte-Urra, I., Warren, H.P.: 2011, *Astrophys. J.* **730,** 37. DOI: 10.1088/0004-637X/730/1/37.

Ulmschneider, P.: 1971, *Astron. Astrophys.* **12**, 297.

Van Doorsselaere, T., Nakariakov, V.M., Verwichte, E.: 2008, *Astrophys. J.* **676**, 73. DOI: 10.1086/587029.

Yang, L., Zhang, L., He, J., Peter, H., Tu, C., Wang, L., *et al.*: 2015, *Astrophys. J.* **15**, 348. DOI: 10.1088/0004-637X/800/2/111.

Zaqarashvili, T.V., Erdélyi, R.: 2009, *Space Sci. Rev.* **149**, 355. DOI: 10.1007/s11214-009-9549-y.